\documentclass[twoside,12pt]{article}  
\usepackage{CJK}   
\usepackage{indentfirst}  
\usepackage{bm}    
\usepackage{graphicx}  
\usepackage{amsmath}  

\begin{document}    



\begin{center}
\LARGE\bf
Discrete Element Crowd Model for Pedestrian Evacuation Through an Exit$\footnote{The study is supported
by the NSFC project (71473207,51178445,71103148) and the Fundamental Research Funds for the Central Universities  (2682014CX103).}$   
\end{center}

\begin{center}
\rm Peng Lin$^a$,  \ \ Jian Ma$^b$ $\footnote{Corresponding
author: majian@mail.ustc.edu.cn}$\ and \ Siuming Lo$^c$
\end{center}

\begin{center}
\begin{small} \sl
${}^{\rm a)}$Department of Fire Safety Engineering, Soutwest Jiaotong University, Chengdu, PR China, drag76@163.com \\   
${}^{\rm b)}$School of Transportation and Logistics, National United Engineering Laboratory of Integrated and Intelligent Transportation, Soutwest Jiaotong University, Chengdu, PR China, majian@mail.ustc.edu.cn \\   
${}^{\rm c)}$Department of Civil and Architectural Engineering, City Univeristy of Hong Kong, Tat Chee Avenue,Kowloon, Hong Kong \\   
\end{small}
\end{center}

\begin{center}
\footnotesize (Received X XX XXXX; revised manuscript received X XX XXXX)
\end{center}

\vspace*{2mm}

\begin{center}
\begin{minipage}{15.5cm}
\parindent 20pt\small
A series of accidents caused by crowd within the last decades evoked a lot of scientific interest in modeling the movement of pedestrian crowds. Based on discrete element method, a granular dynamic model, in which human body is simplified as self-driven sphere, is proposed to simulate the characteristics of crowd flow through an exit. In this model, the repulsive force among people is considered to have an anisotropic feature, and the physical contact force due to body deformation is quantified by the Hertz contact model. The movement of human body is simulated by applying the second Newton's law. The crowd flow through an exit at different desired velocities is studied and simulation results indicated that crowd flow exhibits three distinct states, i.e., smooth state, transition state and phase separation state. In the simulation, clogging phenomenon occurs more easily when the velocity of desired is high and the exit may as a result be totally blocked at a desired velocity of 1.6m/s or above, leading to faster-to-frozen effect.
\end{minipage}
\end{center}

\begin{center}
\begin{minipage}{15.5cm}
\begin{minipage}[t]{2.3cm}{\bf Keywords:}\end{minipage}
\begin{minipage}[t]{13.1cm}
crowd evacuation, discrete element method, anisotropic social force, contact force, desired velocity
\end{minipage}\par\vglue8pt
{\bf PACC: }
45.70.Mg, 05.65.+b, 07.05.Tp
\end{minipage}
\end{center}

\section{Introduction}  
Evacuation system of buildings provides the last defense line for occupants within fire emergency. In the last decades, the evacuation becomes more and more challenging for modern buildings, which are characterized as increase in size, height, occupancy as well as complexity of layout and evacuation routes. Usually, modern building would accommodate a large number of people. When huge amount of people gathering together, the over-congested crowd substantially reduces the exit flow rate and may trigger potential disasters such as trampling and stampede, which resulted in many deaths and injuries in the past decades \cite{1}. Building codes in different countries worldwide stipulate the requirements on the design of egress system, whose performance in real fire scenario is difficult to assess. Design of egress system for complex buildings is a formidable task. With the advancement of simulation techniques and the power of computers, researchers have developed dozens of evacuation models, such as Exodus \cite{2}, SGEM \cite{3}, SFF \cite{Wang2015}, etc., to evaluate the effectiveness of the escape system or evacuation plan in buildings.

These models can deepen our understanding in and may help in preventing accidents due to overcrowding. Thus this research topic has attracted attentions from physicists and sociologists in recent years. Generally speaking, pedestrian movement models can be classified into different categories such as macroscopic continuum models \cite{4, 5}, force-based models \cite{6, 7,Yue2015}, and cellular automaton (CA) models \cite{8, Ma2010a,Ding2015}. Macroscopic models describe the pedestrian movement at a high level of aggregation as fluid flow \cite{4, 5}, which can optimize the overall evacuation network \cite{9}, whilst microscopic models describe the behavior of each individual and their interactions in detail. The movement of people in microscopic models is simplified as a kind of self-driven particle \cite{6, 7, 10,11,12} in which the driving force is not from external, but from each particle itself. Helbing et al. proposed a pedestrian model to quantitatively explain the behavior of panic individuals \cite{6,11,12}. They created a generalized force model using Newton's formula in which force equals pedestrian mass multiplied by its corresponding acceleration. They found "faster is slower effect" in panic evacuation. Over several decades, studies specifically looking at panic behavior in fires have consistently shown that non-adaptive and irrational behaviors are actually a rare occurrence \cite{13,14,15}. After study on the crowd incident in the love parade on 24th July 2010, in Duisberg, Germany, Helbing et al. \cite{16} concluded that there was no sign of panic in the crowd flow during the incident. A special crowd movement pattern, i.e., crowd turbulence, as found in Ref\cite{Ma2013}, should be the cause of the 21 lives in the tragedy. In Ref\cite{7,10}, based on suitable video recordings of interactive pedestrian motion and improved tracking software, the authors proposed  an evolutionary optimization algorithm to determine optimal parameter specifications for the social force model. The calibrated model is then applied in exploring large-scale pedestrian and evacuation simulations including pilgrimage and urban environments. They found the phenomena of intermittent flows and stop-and-go wave in crowd flow at extreme high densities \cite{Yu2007}. Stop-and-go wave is an onset followed by a highly irregular motion of crowd, i.e., the occurrence of crowd turbulence \cite{16,Ma2013,Yu2007}. This crowd dynamics is dangerous and it may cause people fall down. It is noted that the reaction of pedestrians to what happens in front of them is much stronger than to what happens behind them and a centrifugal social force model \cite{17,18} was proposed by taking into account the distance between pedestrians as well as their relative velocities. Recently, researchers modified social force model and analyzed faster-is-slower phenomenon by performing animals experiments \cite{19,Soria2012}. The social-force model has also been used to investigate the effect of complex building architecture on the uncoordinated crowd motion during urgent evacuation \cite{20}. In particular, how the room door size, the size of the main exit, the desired speed and the friction coefficient affect the evacuation time and under what circumstances the evacuation efficiency improves.

Discrete element method (DEM) is a numerical method for computing the motion and effect of a large number of particles \cite{1,21,22,hu2013}. The forces acting on each particle are calculated and force balance is integrated explicitly and acceleration velocity, velocity and the coordinate at each time step is deduced accordingly by applying the second Newton's law. Normally, soft-sphere is used where particles are allowed to slightly overlap. Lin et al. \cite{1} proposed a granular dynamic method to study the egress pattern of evacuees in a densely populated enclosed space. A DEM technique based crowd dynamics model, i.e., CrowdDMX had also been developed by Langston \cite{23}. They further investigated the behaviour of subgroups in crowd dynamics \cite{24}. An agent-based algorithm based DEM algorithm has been incorporated in FDS-Evac \cite{25} and it was further extended to model agents' behaviour in counter-flow situations \cite{26}. However, the faster-is-slower effect, as observed in Helbing's original social force model, is not reported in the CrowdDMX and FDS-Evac despite both the models had adopted social force model as a basic model.

The objective of this paper is to study how the desired velocity affects the throughput of the exit by using the Discrete Element Method. The paper is structured as follows. In section 2, we firstly introduce the original social force model, then discuss its limitations and at last make some revisions accordingly. We will detail the formulation of the proposed model based on theory of granular dynamic. In the 3rd section, parametric study will be conducted. The characteristics of crowd flow at different desired velocity and the flow rate out of the exit are then investigated. The final section presents the conclusions.

\section{Discrete Element Crowd movement model}  
\subsection{Social force model}
In the original social force model \cite{6, 11, 12}, the movement of people is quantified by the following equations,
\begin{equation}
{m_i}\frac{{{\rm{d}}{v_i}}}{{{\rm{d}}t}} = {m_i}\frac{{v_i^0e_i^0(t) - {v_i}(t)}}{\varepsilon } + \mathop \sum \limits_{j( \ne i)} {f_{ij}} + \mathop \sum \limits_w {f_{iw}}
\end{equation}
where ${m_i}$ is the mass of pedestrian $i$, ${v_i}(t)$ is the velocity of pedestrian $i$ at time $t$, ${v_i}^0$ is the desired velocity and $e_i^0(t)$ is the desired direction of movement. $\varepsilon$ is the reaction time and  $\varepsilon$ =0.5s is suggested \cite{Ma2010}. $\mathop \sum \limits_w {f_{iw}}$ is the sum of the contact forces with walls or boundaries. $\mathop \sum \limits_{j( \ne i)} {f_{ij}}$  is the sum force among people and it include two parts. One part is the social force among people to maintain a comfortable distance and the other part is the contacted force due to physical body deformation. The social force is the repulsive force as represented as ${f_{ij}}=A exp[(r_{ij} - d_{ij})/B]$, $d_{ij}$ is the separation distance of two persons and ${r_{ij}} = {r_i} + {r_j}$ is the sums of the radius of the two persons. The parameters $A$ and $B$ are introduced to describe the strength and spatial extent of the force, respectively. $A=2,000N$ and $B=0.08m$ was proposed. The contact force includes the normal contact force $F_n$  and the tangential contact force $F_{\tau}$, and they can be represented as $F_n={k_n}({{r_{ij}} - {d_{ij}}})\overrightarrow{n}$ and $F_{\tau}={k_\tau }\left( {{r_{ij}} - {d_{{\rm{i}}j}}} \right)\Delta {v_\tau }\overrightarrow{\tau}$, respectively.  $\overrightarrow{n}$ is in the normal direction and $\overrightarrow{\tau}$ is in the tangential direction. $\Delta {v_\tau }$ is the tangential velocity difference, $k_n$ and $k_\tau$ are normal contact coefficient and tangential contact coefficient respectively. $k_n$=1.2 $\times$ $10^5$ $kg\cdot s^{-2}$ and $k_\tau$ =2.4 $\times$ $10^5$ $kg\cdot m^{-1}s^{-1}$ are proposed.

\subsection{Discussion on the social force model}
In social force model, a person receives repulsive forces from all people around him so as to keep a comfortable distance to these people. That is to say, distance is an important factor in social force model. While in Centrifugal Force Model \cite{17, 18}, another important factor, the relative velocity was also taken into account. Recently, it was found that the reaction of pedestrians to what happens in front of them is much stronger than to what happens behind them \cite{7}. Thus, a so-called anisotropic social force model is proposed in Ref\cite{24}. In this model,
$f_{_i}^s =A exp[({r_{ij}} - {d_{ij}})/B]\times[\lambda_i+(1-\lambda_i)\times\frac{1+\cos \phi_{ij}}{2}]$
where $\phi_{ij}$ is the angel between the direction of the person $i$ feeling the force and the direction to the person $j$. The parameter $\lambda_i$ represents the anisotropy of the social force. $\lambda_i=1$ corresponds to the original social force model and $\lambda_i<1$ represents the force in front of the person is larger than the force behind. $A=2,000N$ and $B=0.08m$ are implemented in FDS-Evac and CrowdDMX. $A=2,000N$ denotes that the repulsive force is $2,000N$ when two persons contacts but with no physical body deformation, meaning the distance between them is zero. It should be noticed that this force could lead to an acceleration speed of $25m \cdot s^{-2}$ for a $80kg$ person. Even a world $100m$ dash champion like Usain Bolt cannot achieve such a high acceleration speed (his estimated acceleration speed is about $3.09ms^{-2}$ \cite{27}). Therefore, the adopted parameters in the original social force mode and the revised anisotropic social force model are questionable.

Furthermore, the contact force among human body is an important factor to describe the forces among human body when they are physically contacted. In the original social force model, as implemented in FDS-Evac and CrowdDMX, human body is assumed to be an elastic object and the contacted forces are  represented by Hooke's law where the compressive force of human body is linearly dependent on the distance they are pressed. For two spherical particles, the normal push-back force for two overlapping particles is a non-linear function of the overlapped distance but is proportional to the overlapped area as illustrated in Hertz contact model \cite{21,hu2013}. Therefore, the contact force in social force model is over-simplified.

\subsection{Discrete element crowd model}
In DEM simulation, a granular material is modeled based on a finite number of discrete, semi-rigid spherical interacting by means of contact or non-contact forces, and the translational and rotational motions of every single particle in a considered system are described by Newton's laws of motion \cite{28}. In situations such as massive crowd evacuation, each individual does whatever he can to escape, e.g., rushing, pushing, or twisting, to get out the exit as quickly as possible. As a result, translational and rotational movements should both be considered. For simplicity, spherical particles are used in the present study is limited to represent granular pedestrian system.

For the translational movement, we assume that the movement of people is restricted in two dimensional space, $xoy$, while the position in $z$ direction keeps constant. The basic translational motion equations of human body are similar to the ones used in social force, as detailed in former sections. We noticed the animal dynamics reported in Ref\cite{19} and as a consequence assume that the people in front of a person $i$ pose a much higher social impact on him than the persons behind him when there is no physical body contact. That is to say, the social force has an anisotropic feature, as shown in Fig.1. The anisotropic social force $f_{i}^s$ is expressed as,
\begin{equation}
f_{i}^s =A \times exp[({r_{ij}} - {d_{ij}})/B] \times [1+ \varphi \cos(\beta_{ij})]
\end{equation}
where the parameters $A$ is anisotropic social force when the separation distance is zero and $B$ is a parameter related to variation of anisotropic force against separation distance. $\beta_{ij}$ is the angel between the direction of the person $i$ feeling the force and the direction to the person $j$. the parameter $\varphi$ represents the anisotropy of the social force. $\varphi<1$ represents the force in front of the person is larger than the force behind.

\vspace*{4mm} \centerline{\includegraphics[viewport = 0 340 340 540, scale=0.9]{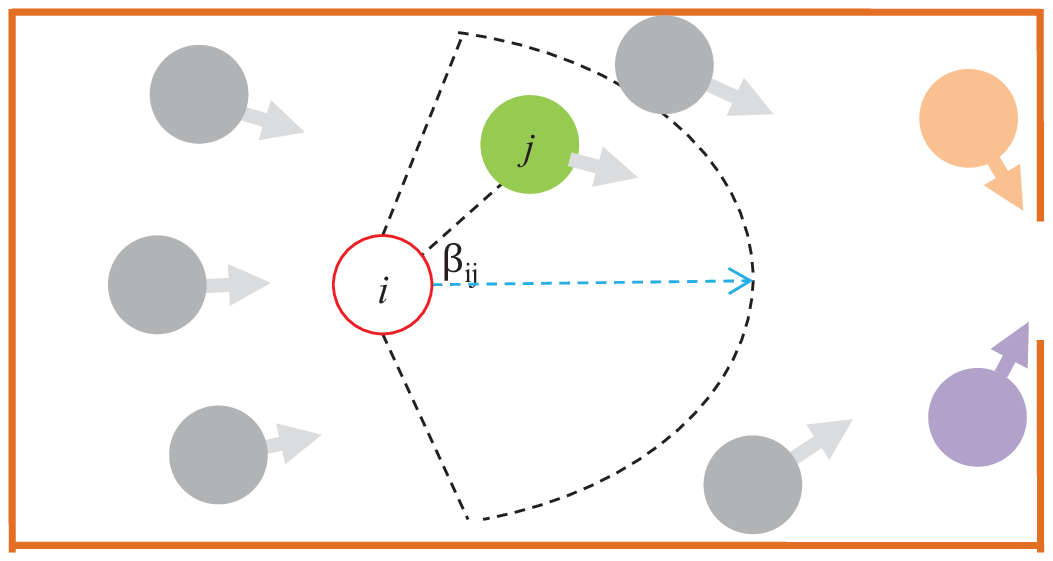}}
\begin{center}
\parbox{15.5cm}{\small{\bf Fig.1.} Scheme of anisotropy social forces posed on a person. }
\end{center}
\vspace*{4mm}

When there is physical body contacts, the contact force is an important factor to describe the forces among human bodies. The mostly used contact model is the Hertz contact model \cite{21} as shown schematically in Fig.2. The normal push-back force for two overlapping particles is proportional to the area of overlap of the two particles, and is thus a non-linear function of overlapped distance. As shown in Fig.2, person $i$ is in contact with person $j$, the normal contract force $F_n$, acting on person $j$, is given by the sum of the normal spring force and normal damping force as,
\begin{equation}
 F_n=\alpha (k_n \delta_n - m_{ij} \gamma_n \Delta v_n)
\end{equation}
The tangential component of the contact force $F_\tau$, is similarly given by the sums of the tangential spring force and the tangential damping force as,
\begin{equation}
 F_\tau=\alpha (k_\tau \delta_\tau - m_{ij} \gamma_\tau \Delta v_\tau)
\end{equation}
where $\alpha=\sqrt{\delta(r_ir_j)/(r_j+r_i )}$, $\delta$ is the overlap distance of two person if they are physically contacted, i.e., $r_{ij} > d_{ij}$  and $\delta=r_{ij} - d_{ij}$ , $k_n$ is the elastic constant for normal contact, $k_n = 2 E/3(1+ \sigma)(1- \sigma)$, in which $\sigma$ is the Poisson ratio, $E$ is the Young's modulus.  $k_\tau$ is the elastic constant for tangential contact, $k_\tau = 2 E/(1+ \sigma) (2- \sigma)$. $\gamma_n$ is the damping constant for normal contact,  $\gamma_\tau$ is the viscoelastic damping constant for tangential contact. $m_{ij}$ is the  effective mass of two persons and $m_{ij}=m_i m_j/(m_i+ m_j)$ , $\delta_\tau$ is the tangential displacement vector between two persons, $v_n$ is velocity different in normal direction $\delta v_\tau$ is tangential component of the relative velocity of the two persons. The magnitude of the tangential force is limited by the Coulomb frictional limit, where the particles begin to slide over each other.

\vspace*{4mm} \centerline{\includegraphics[viewport = 0 300 300 540, scale=1]{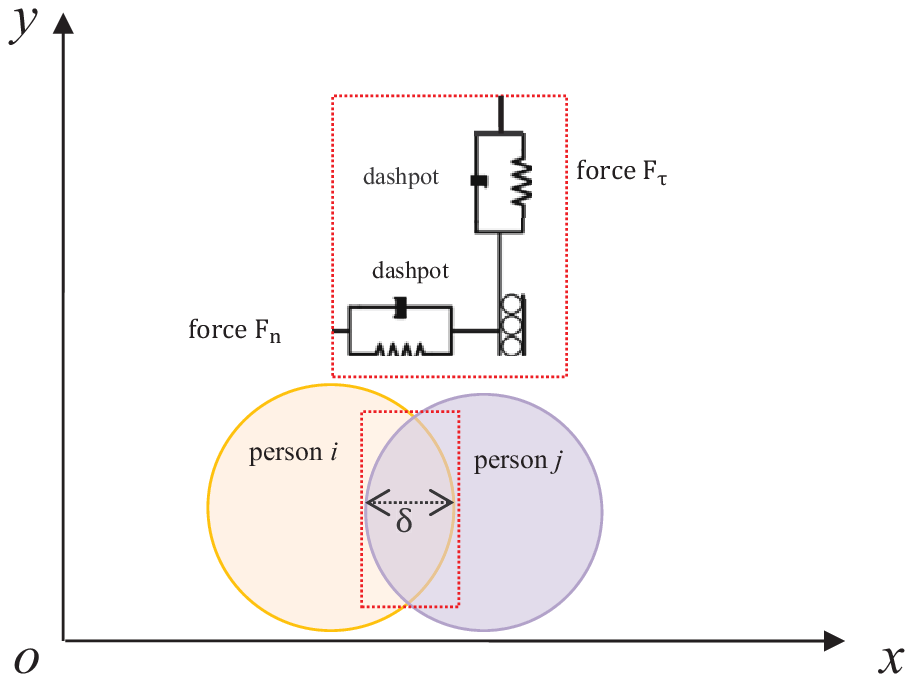} }
\begin{center}
\parbox{15.5cm}{\small{\bf Fig.2.} The Hertz contact model.}
\end{center}
\vspace*{4mm}

The rotational movement of human body is restricted in $z$ direction only and the rotation in direction $x$ and $y$ is set to be zero.
\begin{equation}
{I_i}\frac{{d{\omega _i}}}{{dt}} = \sum {{M_i}}
\end{equation}
where $I_i$ represents the moment of inertia for pedestrian $i$, $\omega_i$ denotes the angular velocity, ${\sum M _i}$ is the total rolling moment, and can be calculated by,
\begin{equation}
{\sum M _i}=\sum_c L_i^c \times F
\end{equation}
 where $L_i^c$ is the vector connecting the center of the mass of particle $i$ with its contact point with particle $j$. $F$ is the sum of all tangential contact forces at contact points. The rolling moment is a result of tangential contact force $f_i$. Considering when a body is standing on an uneven surface in a state of rest, a critical force, i.e., the static friction force $F_s$, must be overcome to set him in motion. This friction force is roughly proportional to the normal force $F_n$, i.e., $F_s= \mu \times F_n$, where $\mu$ is the coefficient of static friction. Kinetic friction $F_k$ is the resisting force which acts on a body after the force of static friction has been overcome. Based on Coulomb's experiment, the kinetic friction is similar to static friction and the coefficient of kinetic fraction is approximately equal to that of static friction, therefore, the dynamic friction among human body is similar to the static friction in the proposed model. That is to say, both static friction and kinetic friction among pedestrians are considered in the present model. If $F_{\tau}> F_s$, the particle can move and the sum of contact forces should be $F=F_{\tau} + F_k$.

The algorithm of translational crowd movement based on the DEM includes the following steps:

Step 1: Creation of the human particles randomly and initialization of all parameters.

Step 2: Calculation of the contact forces, the self-driven force and the social force.

Step 3: Summing up all forces $\sum{\overrightarrow F(t)}$.

Step 4: Based on Taylor expansion and Velocity Verlet integration, the position and velocity at each time step can be expressed as:
\begin{equation}
\vec x(t + \Delta t) = \vec x(t) + \vec v(t)\Delta t + \frac{{\sum {\overrightarrow F (t)} }}{{2m}}\Delta {t^2}
\end{equation}
\begin{equation}
\vec v(t + \Delta t) = \vec v(t) + \frac{{\sum {(\overrightarrow F (t) + \overrightarrow F (t + \Delta t))} }}{{2m}}\Delta t
\end{equation}
This step includes the following four sub-steps,

(a) Using the force and velocity at time step $t$, calculate a half-step velocity $\vec v(t + \frac{{\Delta t}}{2})$ as:
\begin{equation}
\vec v(t + \frac{{\Delta t}}{2}) = \vec v(t) + \frac{{\sum {\overrightarrow F (t)} }}{{2m}}\Delta t
\end{equation}

(b) Calculate the position at time step $t+\Delta t$ as:
\begin{equation}
\vec x(t + \Delta t) = \vec x(t) + \vec v(t + \frac{{\Delta t}}{2})\Delta t
\end{equation}

(c) Using the position $\vec v(t + \frac{{\Delta t}}{2})$, calculate the sum of forces $\vec F(t + \Delta t)$.

(d) Update the half-step velocity $\vec v(t + \frac{{\Delta t}}{2})$ to the full step velocity $\vec v(t + \Delta t)$ as:
\begin{equation}
\vec v(t + \Delta t) = \vec v(t + \frac{{\Delta t}}{2}) + \frac{{\sum {\overrightarrow F (t + \Delta t)} }}{{2m}}\Delta t
\end{equation}

Step 5: Continue the steps from 2-5 until the end of the simulation.

Similarly, the rotational crowd movement can also be realized following similar steps by replacing force, location, speed, acceleration items with moment, orientation, angular speed and angular acceleration items respectively.

\section{Simulation results and discussions}
A number of simulations have been performed for the evacuation of people through an exit. The room is $15m\times15m$ and has an outlet of $1m$ wide as shown in Fig.3. The coordinate of the room is $x\in[0,15]$ in $x$ direction, $y\in[0,15]$ in $y$ direction and $z\in[0,0.38]$ in $z$ direction.  An outlet of $1m$ wide is at center, i.e., $y\in[7,8]$. The diameter of human body is $0.6\pm0.1m$. At the beginning of the simulation, people are randomly distributed in the room and then they will initiate the movement by the self-driven force at their desired velocity. The key input parameters are shown in Table 1. It should be admitted that the present model introduced in parameters more than social force model, thus due to the lack of experimental data, calibration of these parameters need further research work. Thus the model parameters are referenced from data in granular flow theory considering pedestrian and granular flow similarities.

\vspace*{4mm} \centerline{\includegraphics[viewport=0 370 190 540, scale=1]{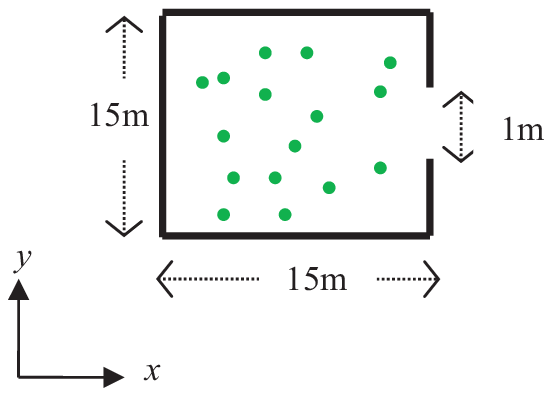} }
\begin{center}
\parbox{15.5cm}{\small{\bf Fig.3.} The schematic layout of the crowd flow through an exit.}
\end{center}
\vspace*{4mm}

\begin{center}
\tabcolsep=15pt  
\small
\renewcommand\arraystretch{1.2}  
\begin{minipage}{15.5cm}{
\small{\bf Table 1.}  Key input parameters in the simulations.}
\end{minipage}
\vglue5pt
\begin{tabular}{| c | c |  }  
\hline 
 {Parameter} & {Value}  \\     
 \hline
  {Number of persons ($N$)} & {100 $\sim$ 300}  \\   
  {Radius of person ($m$)} & {0.25 $\sim$ 0.35} \\   
  {Mass of persons ($kg$)} & {60$\sim$80} \\   
  {Poisson ratio} & {0.25} \\   
  {Normal damping constant} & {0.8} \\   
  {Tangential damping constant } & {0.8} \\   
  {Young's modulus ($MPa$)} & {5} \\   
  {Friction coefficient} & {0.5} \\   
  {Time step length ($second$)} & {0.001} \\   
  {Net simulation time ($minutes$)} & {10} \\   
  {$A$} & {200$\sim$500} \\   
  {$B(m)$} & {0.05} \\   
  {VOD(velocity of desired)($m/s$)} & {1} \\   
\hline
\end{tabular}
\end{center}

\subsection{Parametric analysis}
Firstly, we simulate an evacuation scenario of 100 people getting out a 1-m-wide door at a desired velocity of $1m/s$ and simulation results are shown in Fig.4. With $A=200N$, the leaving time of 100 persons is around 170 seconds, equivalent to a flow rate of $0.6 person\cdot s^{-1}\cdot m^{-1}$. With $A=300N$, the leaving time is around 120 seconds, equivalent to a flow rate of 0.833 $person\cdot s^{-1}\cdot m^{-1}$, which is basically consistent with the flow rate of 0.92 $person\cdot s^{-1}\cdot m^{-1}$ as observed by Fruin in arena evacuation \cite{30}. Further increase the value of $A$ from $300N$ to $500N$ barely affects the flow rate out the exit. Therefore, $A=300N, B=0.05m, \varphi=0.3$ are implemented in subsequent simulations and as a result the repulsive force is $300N$ when two persons are at zero distance.

\vspace*{4mm} \centerline{\includegraphics[viewport=10 10 620 470, scale=0.4]{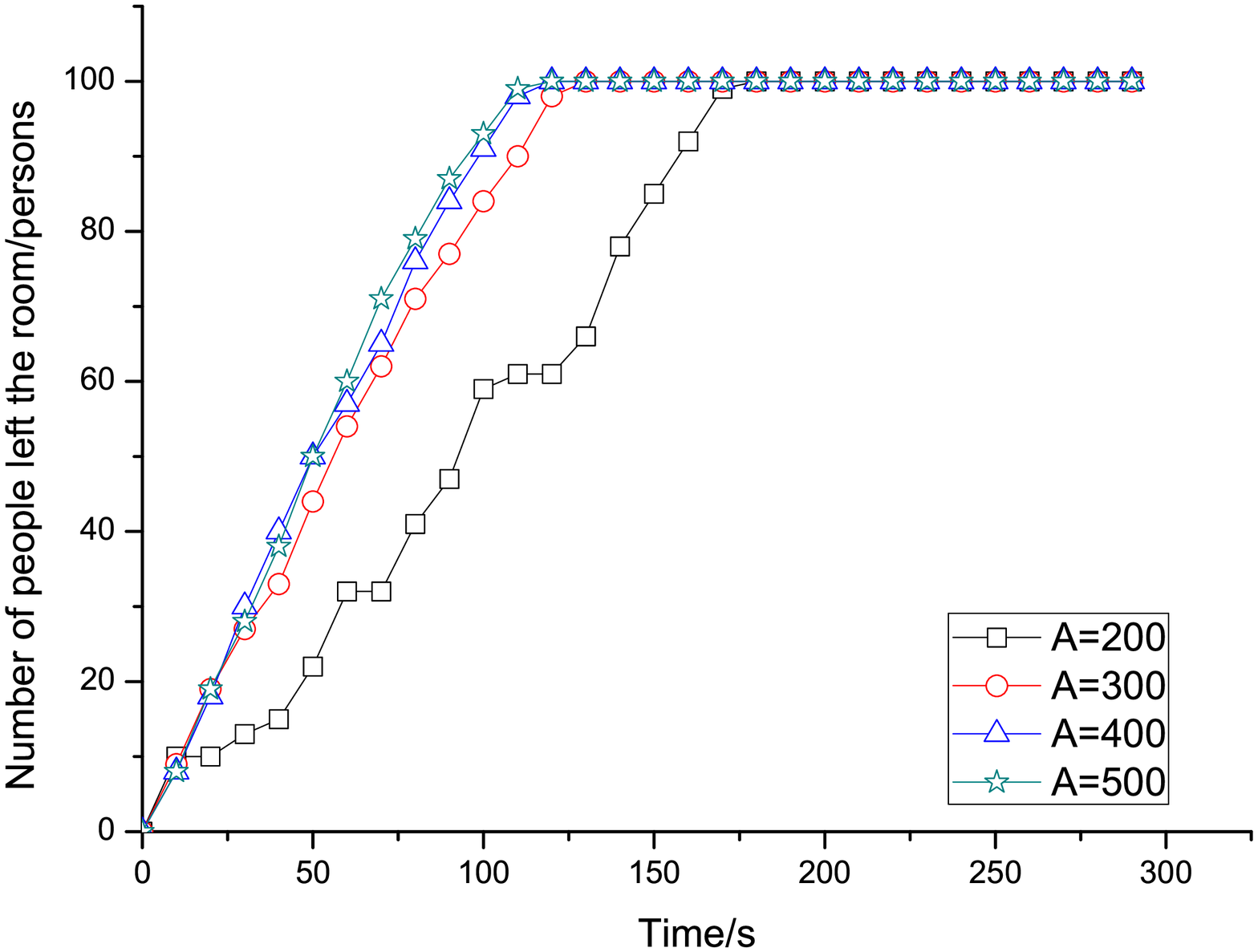} }
\begin{center}
\parbox{15.5cm}{\small{\bf Fig.4.} Evacuation process of 100 people getting out a 1-m-wide door at a desired velocity of 1m/s. }
\end{center}
\vspace*{4mm}

To determine the effect of rotation in the present model, more simulations are conducted by deactivating the rotational Equation (5). Simulation results are shown in Fig.5. From this figure we can see when people moved with a relatively low desired velocity(here we choose 1.0 m/s as a representing case, as shown in Fig.5(a)), the evacuation processes are basically the same with those with rotational equation. However on contrast, when people moved with a higher desired velocity and no rotational movement is considered, as shown in Fig.5(b), the clogging possibility increased. Once clogged, the pedestrian crowds could barely recover to fluent evacuation process. The comparative study showed that the rotational movement of human body, to some extend, can prevent the formation of clog near the exit.

\vspace*{4mm} \centerline{\includegraphics[viewport=0 0 620 470, scale=0.4]{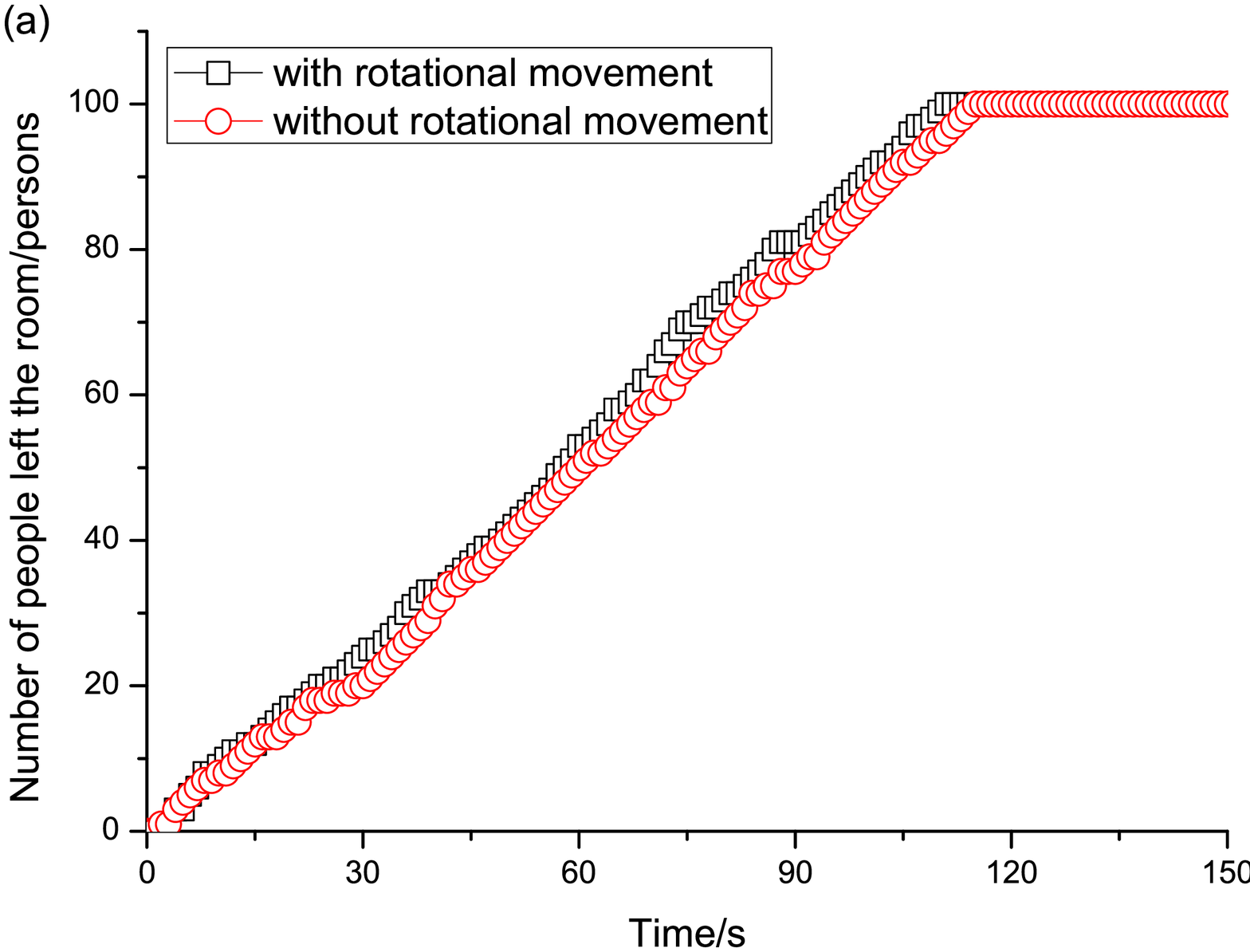}
\includegraphics[viewport=0 0 620 470, scale=0.4]{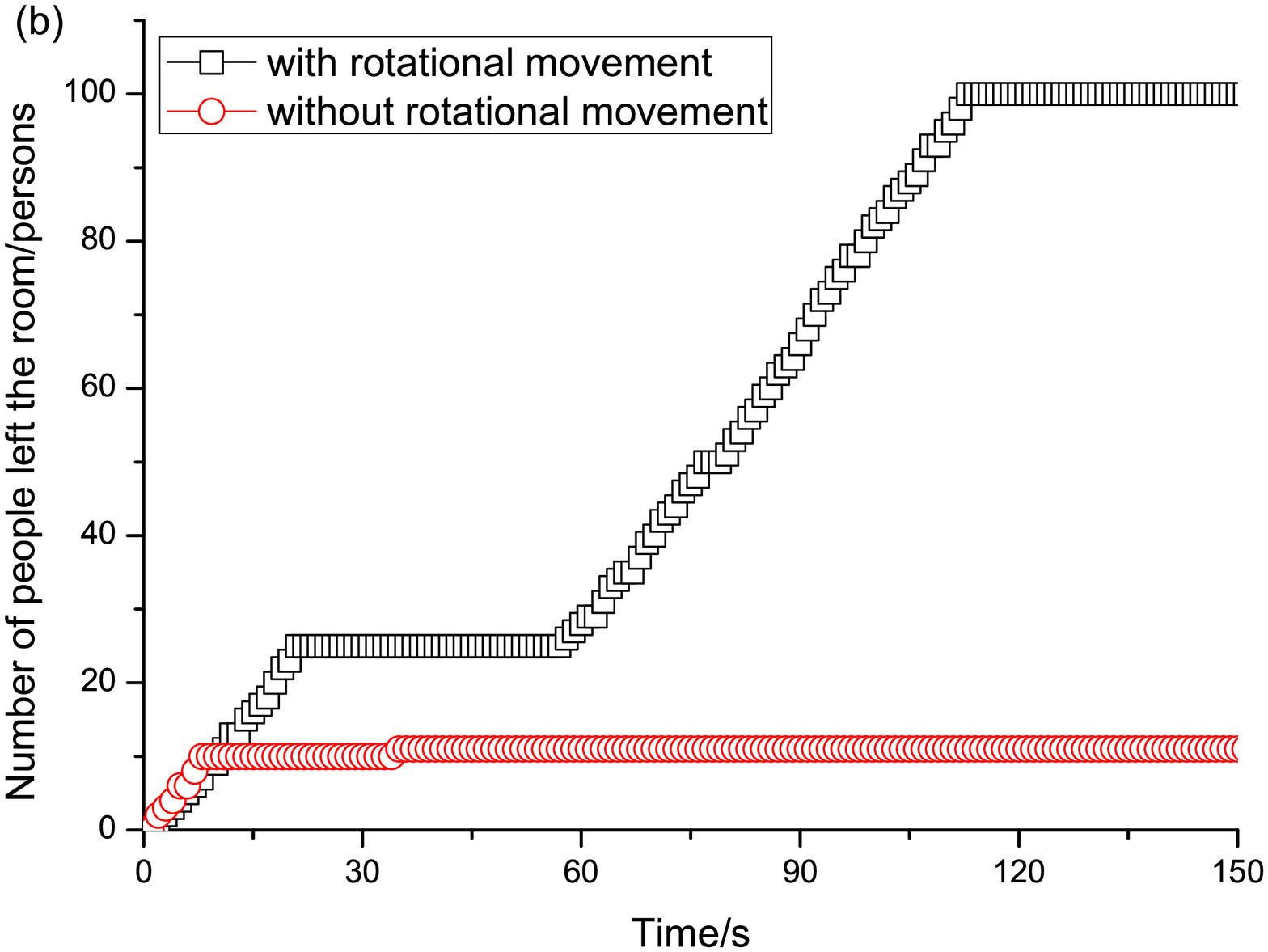} }
\begin{center}
\parbox{15.5cm}{\small{\bf Fig.5.} The evacuation process of 100 people getting out a 1-m-wide door at a desired velocity of (a) 1m/s and (b) 2m/s. }
\end{center}
\vspace*{4mm}

\subsection{Effect of desired velocity}
The clearance time of the occupants within the room is further studied for scenarios with 200 and 300 persons, which are equivalent to an initial density of 0.88 $person/m^2$ and 1.33 $person/m^2$, respectively. Other key input parameters are shown in Table 1.

For 200 people in the room, we show in Fig.6 typical evacuation snapshots. The clearance time at different desired velocities is shown Fig.7. From this figure we can find that the clearance time is approximate 340 seconds at a desired velocity of $0.6m/s$ and it reduced to 210 seconds when the velocity of desired (VOD) increases to $1.0m/s$. With the desired velocity increased from $0.6m/s$ to $1.0m/s$, the clearance time is reduced continuously from 340s to 210s. Further increase in the desired velocity from $1.0m/s$ to $1.8m/s$, the clearance time goes up and down but the average clearance time is close to 220s. With higher desired velocity of $1.8m/s$, the exit is blocked by the evacuees and no one gets out within next 2 minute, i.e., clogging occurs. That means crowd at a desired velocity of $1.8m/s$ or higher could lead to the totally blockage of the exit.

\vspace*{4mm} \centerline{\includegraphics[viewport=0 365 280 540, scale=1]{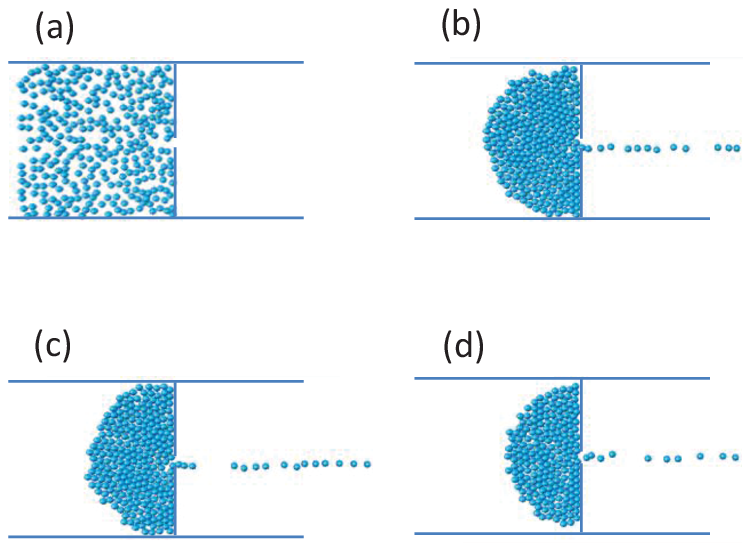}}
\begin{center}
\parbox{15.5cm}{\small{\bf Fig.6.} Snapshots of the crowd evacuation at the time of (a) 0s, (b) 30s, (c) 60s and (d) 120s. }
\end{center}
\vspace*{4mm}

\vspace*{4mm} \centerline{\includegraphics[viewport=0 0 620 470, scale=0.4]{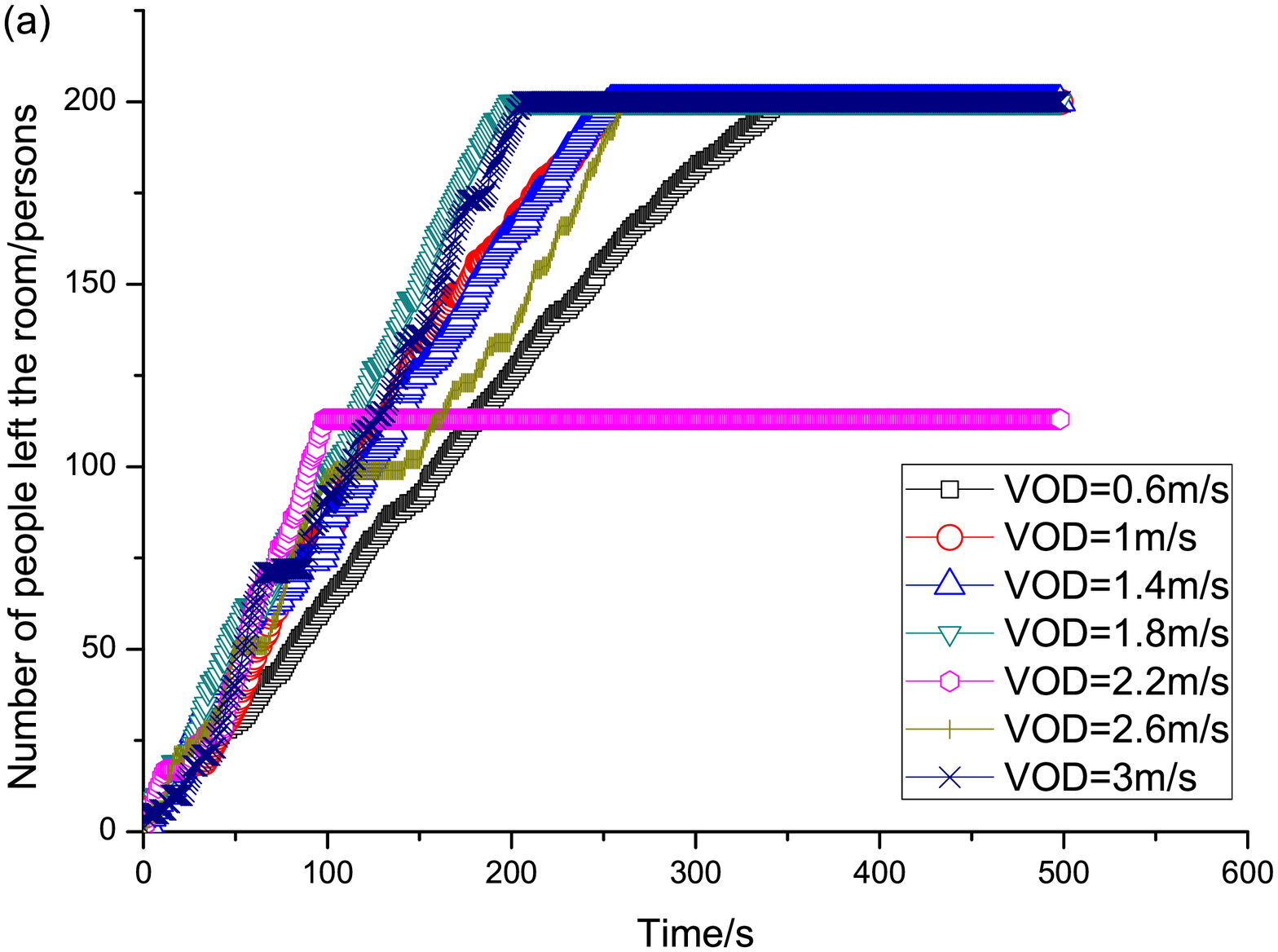}
\includegraphics[viewport=0 0 620 470, scale=0.4]{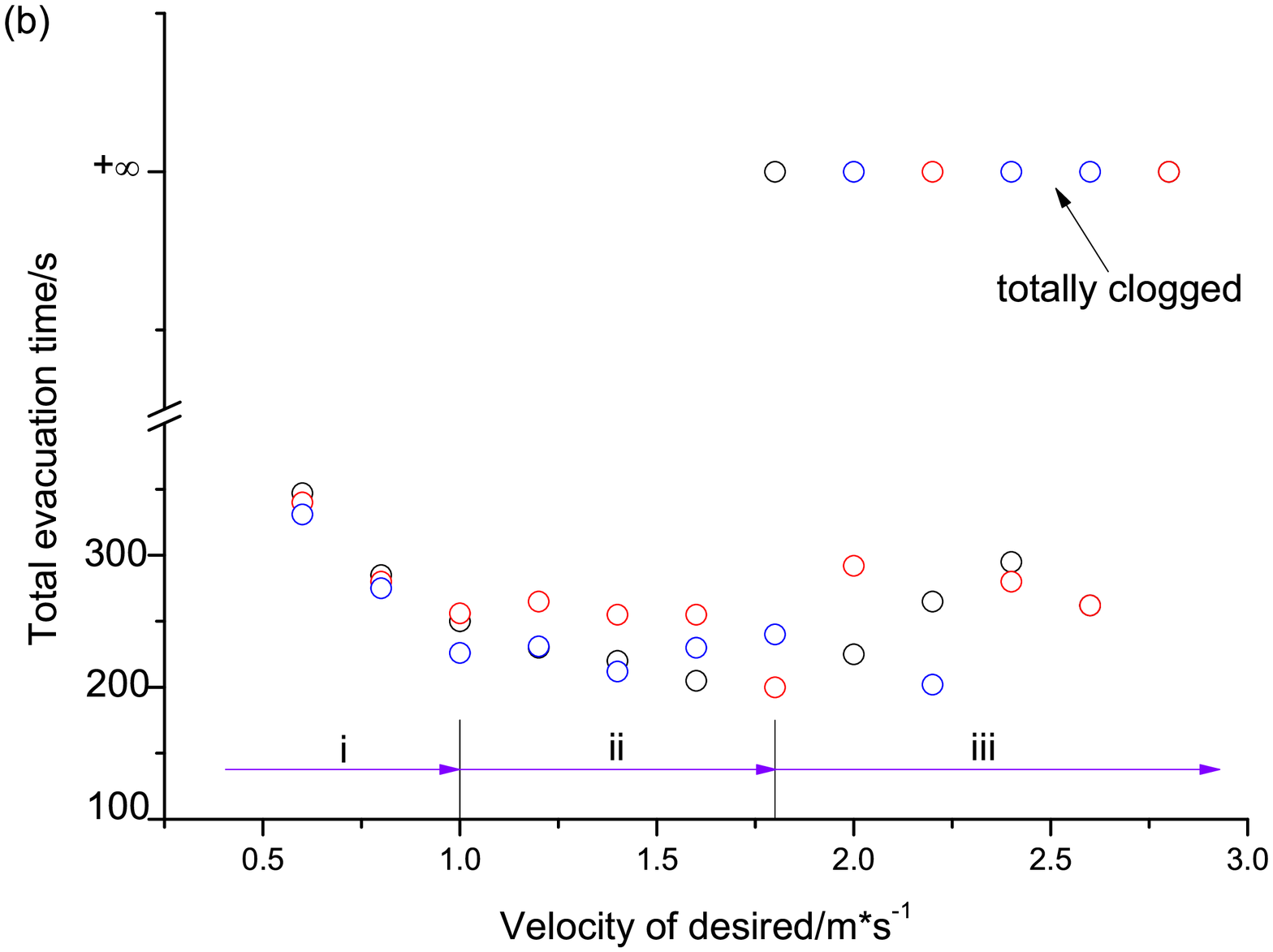}}
\begin{center}
\parbox{15.5cm}{\small{\bf Fig.7.} The leaving process (a) and clearance time (b) of 200 persons at different desired velocity. In (b), different color means different run of the simulation.}
\end{center}
\vspace*{4mm}

Further study with 300 persons is conducted and the results are shown in Fig.8. As can be found in Fig.8, the results are in consistent with that of 200 people. In summary, the relationship between the clearance time and the desired velocity can be classified into three states as follows,

i) Smooth state: in which the clearance time decreases with the increase of the desired velocity from $0.6m/s$ to $1m/s$.

ii) Transition State: in which the clearance time fluctuates when the desired velocity increased from 1.0 m/s to 1.8m/s. The increase in the desired velocity does not decrease the clearance time. No clear trend between the clearance time with and the desired velocity is observed.

iii) Phase separation state: in which the clearance time fluctuates significantly from 80s to infinite (the clearance time is taken as infinite if no one gets out within 2 minutes or people clog at the exit in the end of simulation, i.e.,1000 seconds). That is to say, a bistable region is exhibited, in which either intermittent flow or jammed flow may occur and continue for the entire run. This special state has recently been observed and reported in the experimental granular flow down a chute with two bottlenecks\cite{yang2014}. Total blockage of exit occurred at a desired velocity of greater than $1.8m/s$.

\vspace*{4mm} \centerline{\includegraphics[viewport=0 0 620 470, scale=0.4]{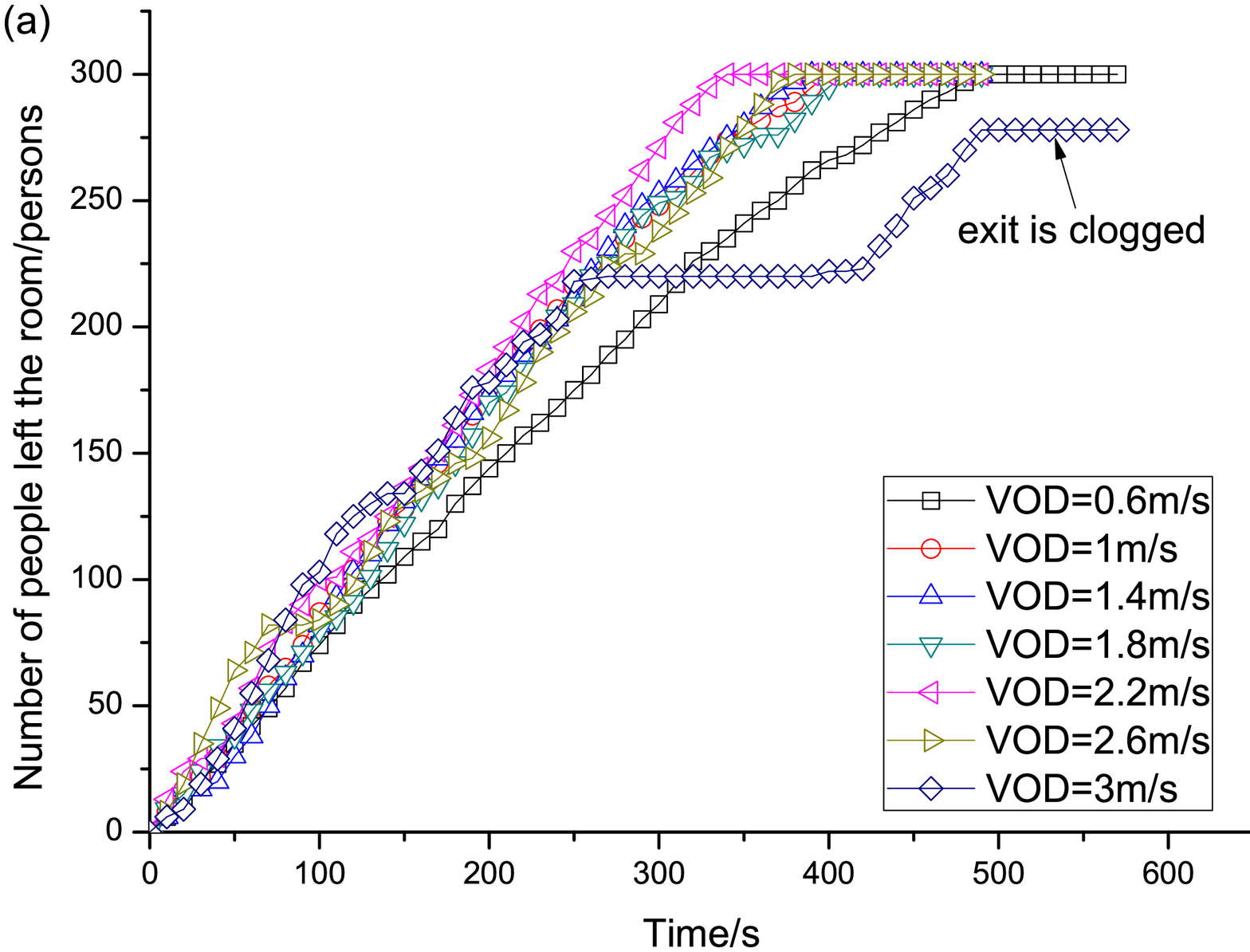}
\includegraphics[viewport=0 0 620 470, scale=0.4]{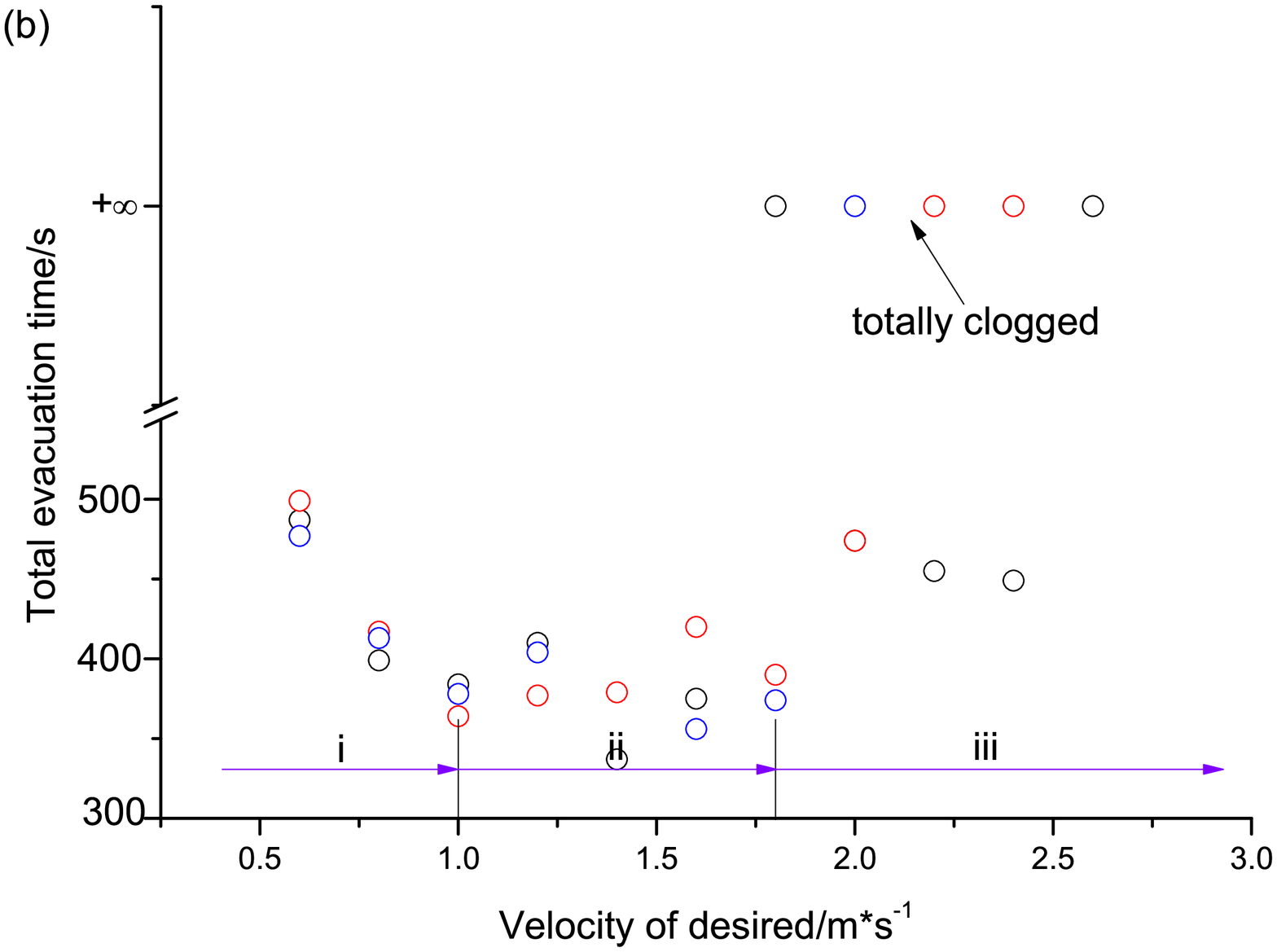}}
\begin{center}
\parbox{15.5cm}{\small{\bf Fig.8.} The leaving process (a) and clearance time (b) of 300 persons at different desired velocity. In (b), different color means different run of the simulation.}
\end{center}
\vspace*{4mm}

\subsection{Exit flow rate}
The exit flow rate is measured at small temporal intervals (10 seconds) to investigate the flow rate at different desired velocity. Examples of typical results are presented in Fig.9. In this figure, the temporary interruptions of the flow are caused by arches that block the exit, while the downward spikes indicates the temporary blockage of exit. When a group of particles flow through a silo in granular flow, the interactions among the particles may lead to the spontaneous development of clogs, as found in experimental granular flow observations\cite{yang2014,Lozano2013,hu2015}. The intermittent flow was also observed in the crowd flow through an exit as shown in Fig.8 and Fig.9(a).

In social force model, we know a special phenomenon named "faster-is-slower" is observed. The quickest leaving time of 200 persons is around 150s at a desired velocity of 1.5m/s and the leaving time increases to 200s at a desired velocity of 5m/s \cite{12}. Here in Fig.9, we can find that for a desired velocity of $0.6m/s$, the flow rate varies from $0.4person/s$ to $0.6person/s$. At a desired velocity of $1 \sim 1.2m/s$, the flow rate varies from $0.3person/s$ to $1.3person/s$. At a desired velocity of $1.6m/s$ or above, the flow rate varies from 0 to $1.6person/s$. Although the overall tendency of the evacuation times is consistent with the "faster-is-slower" phenomenon, we can observe totally jammed phase in our simulation even the desired velocity is not so large.

\vspace*{4mm} \centerline{\includegraphics[viewport=0 0 620 470, scale=0.4]{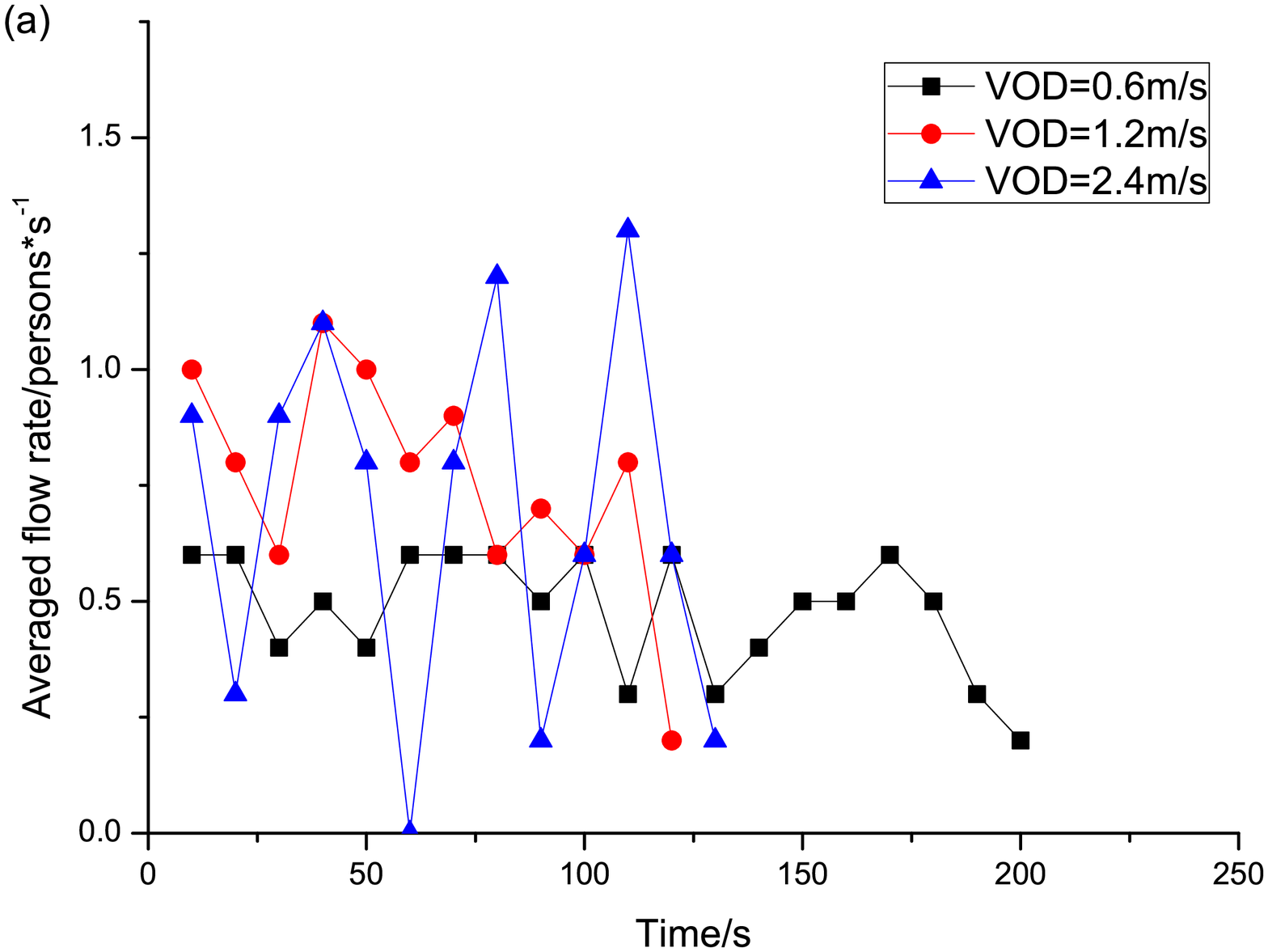}
\includegraphics[viewport=0 0 620 470, scale=0.4]{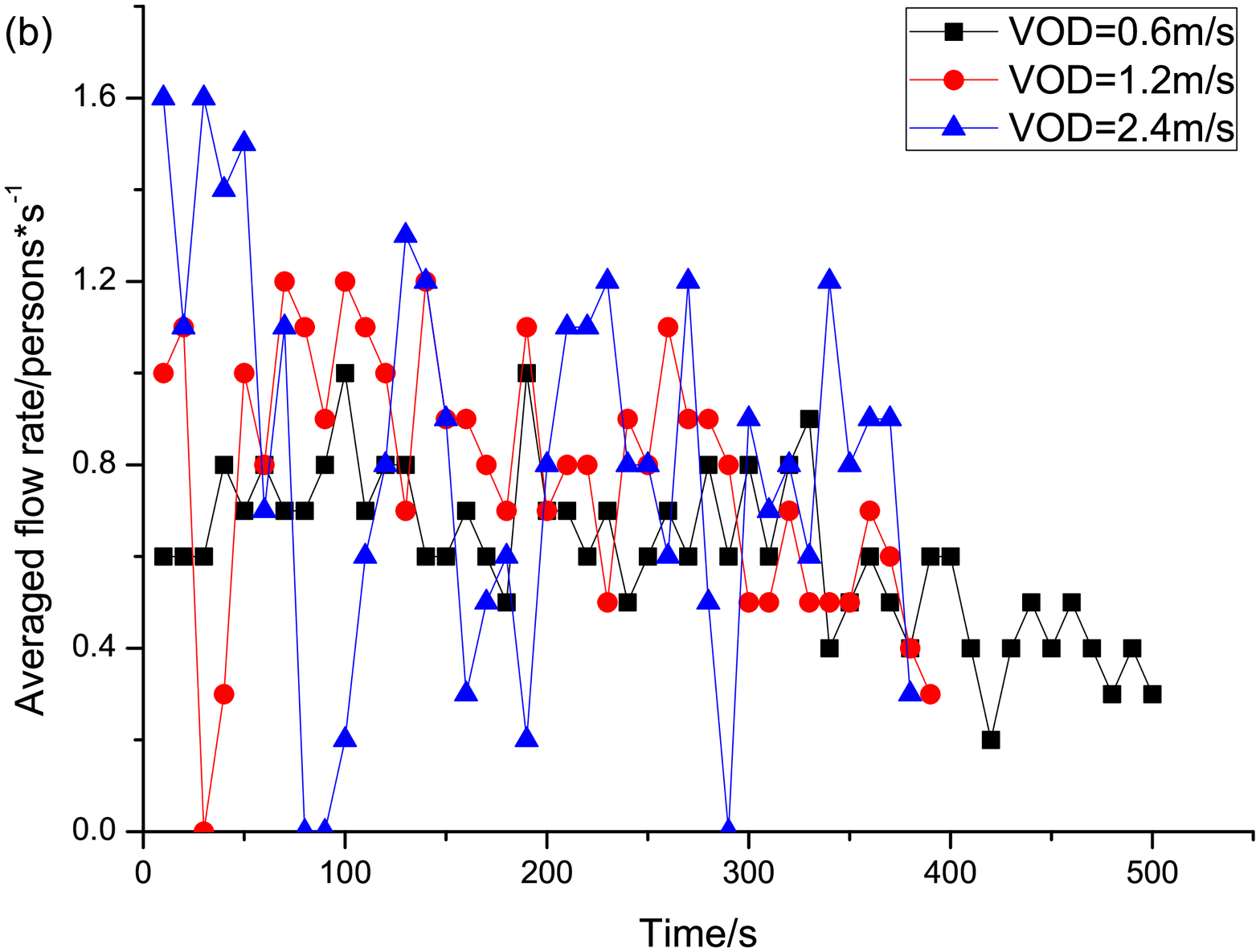}  }
\begin{center}
\parbox{15.5cm}{\small{\bf Fig.9.} The flow rate of 100 people (a) and 300 people (b) getting out the exit at different desired velocities.}
\end{center}
\vspace*{4mm}

The average flow rate of people getting out of the exit is further investigated with different initial number of people to be evacuated. The results are summarized in Fig.10. From this figure we can see that the relationship between the flow rate and the desired velocity can also be classified into three states. i) The smooth state, in which flow rate increases with the increase of the desired velocity from $0.6m/s$ to $1m/s$. ii) The transition state, in which flow rate fluctuates from $0.7person/s$ to $0.95person/s$ with the desired velocity increased from $1.0m/s$ to $1.6m/s$. The impatience of crowd neither reduces the flow rate, nor improves the efficiency of the exit. The flow rate keeps at an average of $0.82person/s$. iii) Phase separation state, in which flow rate fluctuates greatly from $1.22person/s$ to zero when the desired velocity of crowd is greater than $1.6m/s$. The crowd movement at the exit would become frozen if desperate crowd at a desired velocity of greater than $1.6m/s$, which lead to the faster-to-frozen effect. Meanwhile, in Fig.8, phase transition is observed where the system transforms from a state of free movement to a state of totally jam or blockage at exit and the throughput decreases dramatically near the transition state when the degree of impatience increases.

\vspace*{4mm} \centerline{\includegraphics[viewport=0 0 620 470, scale=0.6]{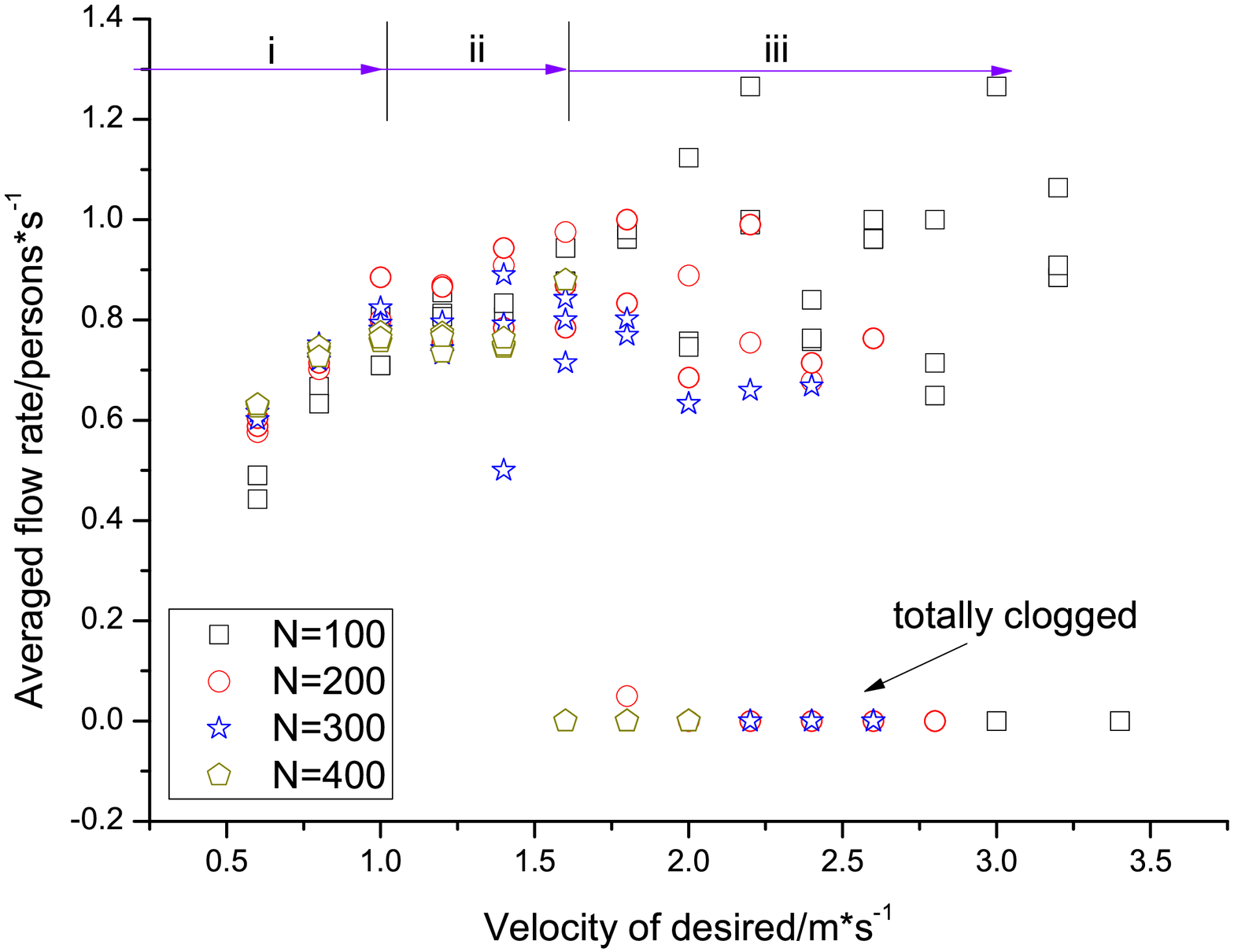} }
\begin{center}
\parbox{15.5cm}{\small{\bf Fig.10.} The flow rate of crowd getting out the exit at different desired velocity.}
\end{center}
\vspace*{4mm}

\section{Concluding remarks}
In the original social force model, the parameters for the social force is questionable and the contact force due to body deformation is over-simplified. Thus in the present study, we built a model for crowd flow through an exit by adopting the theory of granular dynamic. In the proposed model, human body is treated as granular particle and the dynamical movement of people is studied by applying the second Newton's law. The Hertz contact model is incorporated to describe the contact forces among human body when they are physically contacted. The friction force among particles is also considered, and finally a revised anisotropic discrete element crowd model is formulized. Within this model, the effect of velocity of desired (VOD) were investigated. It was assumed that a larger desired velocity means a higher degree of intention people want to get out the room. The clearance time and flow rate of people getting out of a room at different desired velocities were investigated. The relationship between the clearance time/flow rate and the desired velocity exhibits three distinct states as follows:

 i) Smooth state: in which crowds get out in an order manner, i.e., in a normal walking velocity ranged from $0.6m/s$ to $1m/s$ and the flow rate increases with the increase of the desired velocity.

 ii) Transition state, in which crowds get out with desired velocity from 1m/s to 1.6m/s. The flow rate fluctuates slightly from $0.7person/s$ to $0.95person/s$. The increase in the desired velocity adds the fluctuation in the flow rate and the overall average flow rate does not improve much.

 iii) Phase separation state, in which crowds get out with desired velocity of higher than $1.6m/s$. The flow rate fluctuates greatly from $1.2person/s$ to zero and the total frozen of crowd near the exit is observed due to crowd striving to get out as quickly as possible, which results in a dead block.

Our simulations show that the clogging phenomenon occurs more easily at exit if all people still try to get out at a desired velocity irrespective of the net distance/space available. The competitiveness of crowd cannot improve the flow rate but only adds its fluctuation and the over-competitiveness of crowd could lead to the clog at the exit, i.e., faster-to-frozen effect.

It should be noted the current model only considers the rotation of human body due to external contact force. In fact, people are self-driven particles and they can provide internal rotational force so as to twist or rotate their body if they are jammed at the exit. It is believed that the internal rotational force could reduce the clog possibility. It should also be noticed that in the current model, the evacuation process with only one exit was considered, whilst two or more exit is generally provided in the reality. By introducing in novel potential field such as the one proposed in Ref\cite{zhang2012,xiong2011}, multi-exit evacuation will be further studied in the future.

\vspace*{2mm}


\end{document}